\documentstyle[bo99,epsfig]{article}

\title{Synchrotron and Compton Components and their Variability in BL Lac Objects}

\author{P.Giommi $^1$, G. Ghisellini $^2$, P. Padovani $^{3,4}$,
and G. Tagliaferri$^2$}

\affil{1) BeppoSAX Science Data Center, Via Corcolle 19, 00131 Rome, Italy\\
       2) Osservatorio Astronomico di Brera, Via Bianchi 46, 23807 Merate, Italy\\
       3) Space Telescope Science Institute, San Martin Drive, Baltimore, MD 21218 U.S.A. \\
       4) Affiliated to the Astrophysics Division, Space Science Department, European Space Agency}

\begin{document}

\maketitle



\section{Introduction}

BL Lacertae objects are extreme extragalactic sources characterized by 
the emission of strong and rapidly variable nonthermal
radiation over the entire electromagnetic spectrum. Synchrotron emission
followed by inverse Compton scattering in a relativistic beaming scenario
is generally thought to be the mechanism powering these objects
(e.g. Kollgaard 1994., Urry \& Padovani 1995).
BL Lacs can be divided into different subclasses depending on 
their Spectral Energy Distribution (SED), 
namely LBL for objects with the synchrotron emission peaking at
$\nu_{peak}\approx 10^{13-14} Hz $, intermediate objects
($\nu_{peak}\approx 10^{15-16} Hz $) and HBL or high energy peaked BL Lacs
with $\nu_{peak}\approx 10^{17-18}Hz$ (Padovani \& Giommi 1995).
The wide X-ray band pass of the BeppoSAX satellite (Boella et al. 1997)
is well suited for the detailed spectral study of all types of BL Lacs.
In fact, direct measurements of the Compton part of the spectrum have been obtained 
for a number of LBLs (e.g. Padovani et al. 1999), and the very variable
tail of the 
Synchrotron component has been studied in several HBLs (e.g. Pian et al. 1998,
Wolter et al. 1998, Giommi, Padovani \& Perlman  1999, 
Chiappetti et al. 1999).
In the case of the two intermediate BL Lacs S5~0716+714 and ON~231 BeppoSAX 
for the first time was able to detect both spectral components 
within a single instrument (Giommi et al. 1999, Tagliaferri et al. 1999).

The BeppoSAX archive at the Science Data Center (SDC, Giommi \& Fiore 1998)
presently includes over 100 observations of 56 distinct BL Lacs, about  
half of which are already publicly available.
We have started a project to construct the SED of a large number of all 
types of BL Lacs by combining a) public BeppoSAX data (0.1-200 keV);
b) simultaneous optical and radio data when these are available
from monitoring campaigns, or from the University of Michigan Radio
Astronomy Observatory (UMRAO) on-line data base (Aller et al. 1999);
and c) non-simultaneous photometric data form NED.
Here we present the first results of this project.

\begin{figure}[ht]
\centerline{\psfig{file=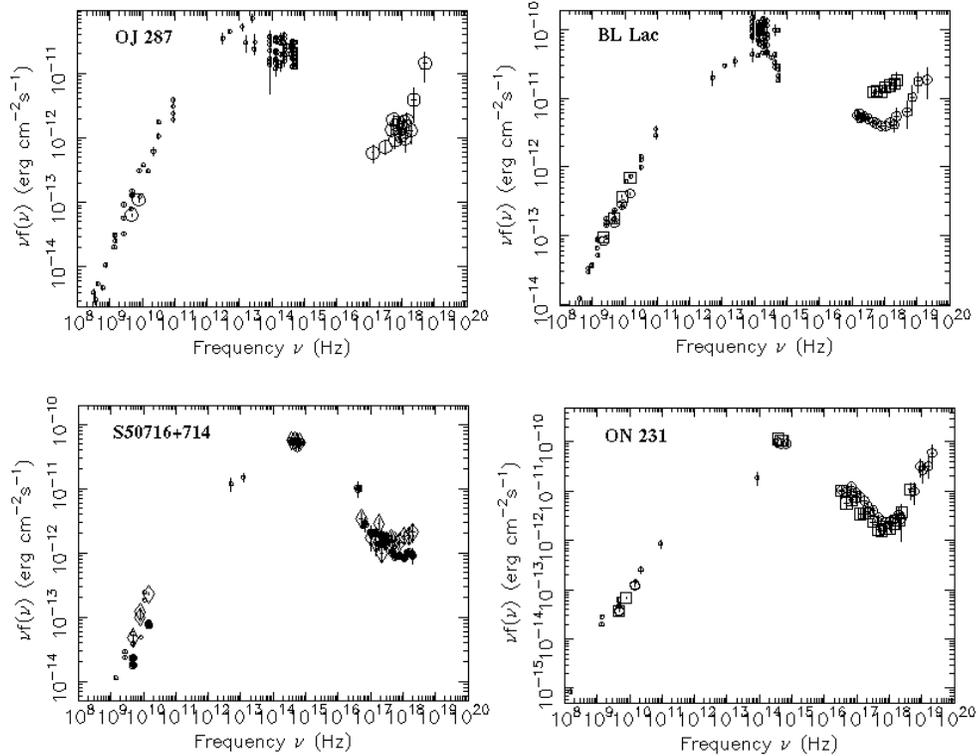, width=13.0cm}}
\caption[]{Spectral energy distribution of four LBL and intermediate 
BL Lacs.}
\end{figure}

\section{Spectral Energy Distributions and variability}

\begin{figure}[ht]
\centerline{\psfig{file=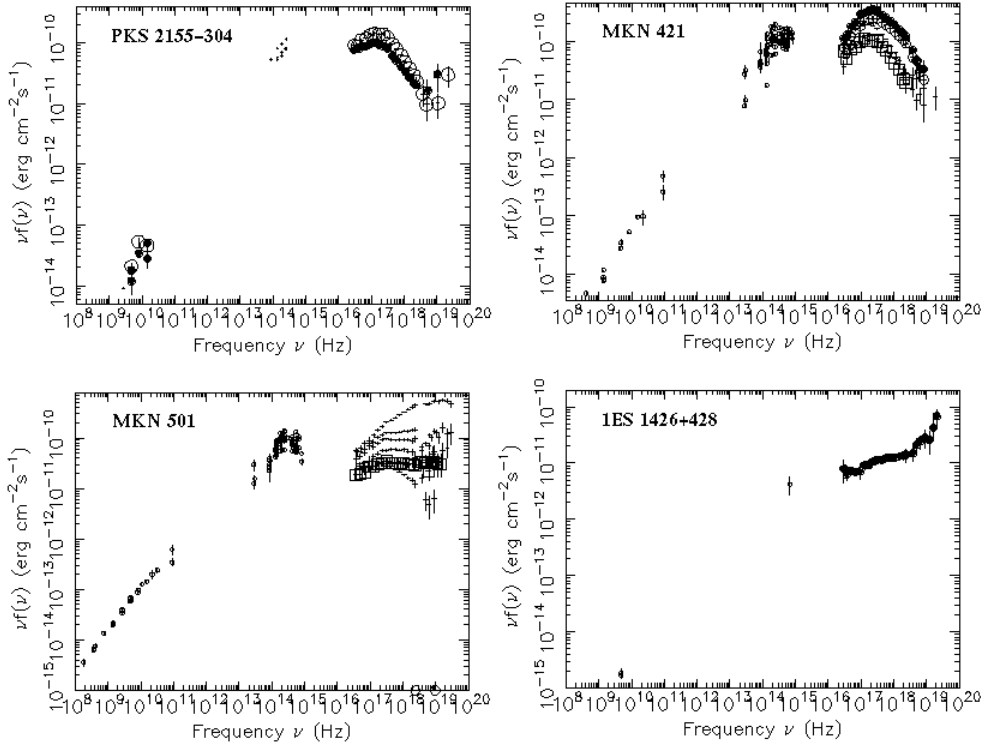, width=13.0cm}}
\caption{Spectral energy distribution of HBL BL Lacs.}
\end{figure}
The SEDs that we have assembled are shown in figure 1 for LBLs and
intermediate objects, and in figure 2 for HBL BL Lacs.
The X-ray part of the plots have been constructed using data from the
LECS, MECS and PDS instruments of the BeppoSAX satellite.
The cleaned and calibrated data files have been taken
from the SDC on-line archive and have been analyzed using the XSPEC package.
Unfolded spectral data have been corrected for low energy absorption
assuming $N_H$ equal to the Galactic value. 
Nearly simultaneous data are plotted with the same symbols used for
the X-ray data. Optical monitoring observations are available 
for S5~0716+714 (Giommi et al. 1999), and ON~231 (Tagliaferri et al. 1999).
Nearly simultaneous radio data from the UMRAO database are
available for several objects. All other (non-simultaneous) data are
plotted as small open circles and are from the photometric data points
provided by NED.
Strong variability at several frequencies is apparent from Figures 1 and 2.
In particular quite spectacular spectral changes are concentrated at or just 
after the synchrotron peak. BeppoSAX observations of intermediate BL Lacs 
clearly show that the soft X-ray synchrotron radiation vary in a different 
way compared to the harder Compton components (Giommi et al. 1999, 
Tagliaferri et al. 1999).
The SED shown here indicate that the variability of the Compton component
may be correlated with radio flux and not with the
optical and soft X-ray synchrotron emission (see figure 1).

\section{Ultra High Energy Synchrotron Peaked BL Lacs (UHBLs) ?}

Figures 1 and 2 clearly show that the peak frequency of the
synchrotron emission ranges from around $10^{13}Hz$ for OJ 287 to
well above $10^{19}Hz$ for 1ES1426+428. Ghisellini (1999) argued that this
trend could continue to much higher energies.
We have thus been searching for BL Lacs with Ultra High
synchrotron peak energy (UHBLs). We have selected candidates
UHBLs from the sample of extreme BL Lacs of the "Sedentary 
Multifrequency Survey" (Giommi, Menna \& Padovani 1999) by looking for 
objects within the error circle of unidentified sources in the 
third EGRET catalog.
One such object is 1RXS J23511.1-14033; its finding chart 
is shown in figure 3 (left). The SED of 1RXS J23511.1-14033, on the right 
part of figure 3, indicates that the
synchrotron emission could reach the gamma ray band. A first BeppoSAX
pointing of this object unfortunately gave inconclusive results since the
observation had to be split into three short exposures and the spectrum 
appears to be variable. Details will be published elsewhere. A second UHBL 
candidate will be observed by BeppoSAX in a few months. If these 
observations will confirm the hypothesis that UHBLs exist, this type
of sources could be the long sought counterpart of many of the
still unidentified high galactic latitude EGRET sources.

\begin{figure}
\centerline{\psfig{file=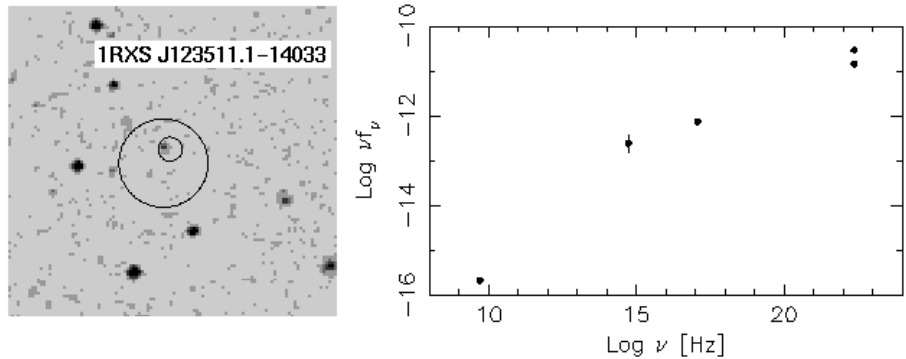, width=13cm}}
\caption[]{Left: the ROSAT and NVSS error circles showing the candidate 
UHBL 1RXS J123511.1-14033. 
Right: the SED of 1RXS J123511.1-14033 if this BL Lac is the correct
counterpart of the EGRET source 2EGJ1233-1407}
\end{figure}



\end{document}